\documentclass[english,aps,prc,superscriptaddress,amsmath,amssymb,showpacs]{revtex4}
\usepackage[T1]{fontenc}
\usepackage[latin9]{inputenc}

\usepackage{bm}\usepackage{amstext}

\usepackage{babel}

\begin{document}

\title{Calculations and observations for a number of $N=Z$ nuclei}

\author{S.J.Q. Robinson}
\author{T. Hoang}
\affiliation{Department of Physics, Millsaps College, Jackson, Mississippi, 39210}

\author{L. Zamick}
\author{A. Escuderos}
\author{Y.Y. Sharon}
\affiliation{Department of Physics and Astronomy, Rutgers University, Piscataway,
New Jersey, 08854}

\begin{abstract}
In this work we look at the low lying nuclear structure of several
$N=Z$ nuclei residing between the doubly magic nuclei $^{40}$Ca and
$^{100}$Sn. Using large shell model codes, we calculate and discuss
the systematics of energies. We show energy levels, $B(E2)$'s, static
quadrupole moments and $g$ factors. In all cases, we compare the results
of two different interactions which yield significantly different occupation
numbers. We compare with the simplest versions of the rotational and
vibrational models. By examining $B(E2)$'s and static quadrupole moments,
we make associations with collective models and find that, in the model
space here considered, $^{88}$Ru is oblate. The quadrupole moment
of $^{92}$Pd is very small, consistent with the vibrational model.
\end{abstract}

\maketitle

\section{Introduction}

In this work we use large scale shell model calculations to study
properties of $N=Z$ even-even nuclei. We consider energy levels, $B(E2)$'s,
static quadrupole moments and magnetic $g$ factors. Many of the quantities
that we calculate have not been measured, especially static quadrupole
moments of high spin states. However, they are useful for seeing how
the shell model stacks out in comparison with collective models.

Large space shell model calculations of energy levels and $B(E2)$'s
in the $f_{7/2}$ shell were performed in the past by Robinson, Escuderos
and Zamick~\cite{sjqr05}. They calculated $B(E2)$'s to high spin states
in $^{44}$Ti and $^{48}$Cr. Such calculations will be also done
here, but instead of calculated transitions from $J$ to $J+2$, we will
reverse and go from $J+2$ to $J$. This makes comparisons with the vibrational
model easier. We will also include heavier nuclei $^{96}$Cd, $^{92}$Pd
and $^{88}$Sr. Indeed this study is in part motivated by the recent
work of Cederwall and collaborators on $^{92}$Pd~\cite{cederwall11}.
They note that the energy levels of $^{92}$Pd are equally spaced
but the $B(E2)$'s are closer to the rotational model.

In the $f$-$p$ region, two interactions are used, gxpf and FPD6; in the
heavier mass nuclei, which will require the $g_{9/2}$ shell, we used
jun45 and jj4b. One of the purposes of this work is to compare the occupancy
numbers with these different interactions, as well as the consequences
of these differences.

It was noted in Ref.~\cite{sjqr05} that the $B(E2)$'s dropped as one
went to the highest spins allowed by the $f$-$p$ model space. As we will
see in the next section, this is quite different from what happens
in the simplest versions of collective models.

\section{Collective Models}

In the rotational model, the formulas for $B(E2,J\rightarrow J-2)$
and $Q(J)$ are related to the intrinsic quadrupole moment $Q_{0}$ as
follows:
\begin{equation}
B(E2,J\rightarrow J-2) = \frac{5}{16\pi} (J 2 K 0 |(J-2) K)^{2}
Q_{0}^{2}
\end{equation}

\begin{equation}
Q(J) = \frac{3K^{2} - J(J+1)}{(J+1) (2J+3)} Q_{0}
\end{equation}

For a $K=0$ band, we also have 
\begin{equation}
B(E2,J\rightarrow J-2) = B(E2, 2\rightarrow 0) 
\frac{15 J(J-1)^{2}}{(2J-2) (2J-1) (2J+1)} \, . 
\end{equation}

In this model the relation between the static quadrupole moment and
$B(E2)$ for $J=2^+$ is
\begin{equation}
Q(2^+) = -2.0256607 \sqrt{B(E2,2\rightarrow0)}
\end{equation}
in the prolate case. For higher values of $J$, we have 
\begin{equation}
Q(J)= 3.5 \frac{J}{2J+3} Q(2^+) \, .
\end{equation}

As $J$ becomes very large, the ratio $B(E2,J\rightarrow J-2)/B(E2,2\rightarrow 0)$
reaches an asymptotic limit of $15/8=1.875$, whilst $Q(J)/Q(2)$ reaches
a limit of 7/4.

In the vibrational model the $B(E2)$ for the yrast sequence $J=0,2,4,6,8$,
etc. is given by
\begin{equation}
B(E2,J+2 \rightarrow J) = \frac{J+2}{ 2} B(E2,2\rightarrow 0) \, ,
\end{equation}
i.e. the $B(E2)$ is proportional to the number of quanta and increases with
$J$. The static quadrupole moment vanishes.

As far as energy levels are concerned, in the simple rotational model
one has a $J(J+1)$ spectrum. For the yrast sequence $J=0,2,4,6$, etc.,
one gets equally spaced levels in the harmonic vibrational model.

In either collective model, the $g$ factors for the states is given
by $Z/A$, which in this work is 0.5 as we are considering $N=Z$ nuclei.

\section{Excitation Energies}

The calculated excitation energies, $B(E2)$'s, $g$ factors and static
quadrupole moments are shown in Tables~\ref{tab:44ti} to \ref{tab:96cd}
for $^{44}$Ti, $^{48}$Cr, $^{52}$Fe, $^{88}$Ru, $^{92}$Pd, and $^{96}$Cd. 
In all cases the energy levels are neither pure rotational or pure vibrational. 
One can say however that they are overall closer to vibrational with deviations
towards the rotational.

For high spins one can get crossovers which lead to long-lived isomeric
states. Experimentally, in $^{52}$Fe the $12^{+}$ state comes below
the $10^{+}$ state. The $12^{+}$ cannot decay via an $E2$ transition
and has a half-life of over 15~minutes. In the rotational model, one
would not get a crossover if both the 10 and 12 were members of a
$K=0$ band. In the large scale shell model, we fail to get the crossover 
with the fpd6 interaction, whereas with gxfp1 the two states are
almost degenerate.

\begin{table}[htb]
\caption{Excitation energies, $B(E2)$'s, $g$-factors, and quadrupole moments of
$^{44}$Ti using the gxpf1 (FPD6) interaction.}
\label{tab:44ti} 
\begin{ruledtabular}
\begin{tabular}{cccccc}
Yrast state & Theor. energy & Exp. energy & $B(E2)$ $\downarrow$  & $g$-factor  & Quadrupole moment \\
\hline 
2$_{1}^{+}$  & 1.408 (1.300) & 1.083 & 103 (139.8)  & 0.546 (0.514)  & -5.1 (-21.7) \\
4$_{1}^{+}$  & 2.552 (2.498) & 2.454 & 133.1 (190.4)  & 0.538 (0.515)  & -16.4 (-29.0) \\
6$_{1}^{+}$  & 3.295 (3.775) & 4.015 & 103.2 (160.9)  & 0.528 (0.519)  & -30.7 (-33.4) \\
8$_{1}^{+}$  & 5.521 (6.248) & 6.508 & 70.5 (111.9)  & 0.551 (0.540)  & -19.5 (-27.1)\\
10$_{1}^{+}$  & 6.678 (7.613) & 7.671 & 92.3 (109.4)  & 0.549 (0.546)  & -22.7 (-25.7) \\
12$_{1}^{+}$  & 7.085 (8.312) & 8.039 & 53.8 (63.3)  & 0.549 (0.549)  & -28.3 (-28.5) \\
\end{tabular}
\end{ruledtabular} 
\end{table}

\begin{table}[htb]
\caption{Excitation energies, $B(E2)$'s, $g$-factors, and quadrupole moments of
$^{48}$Cr using the gxpf1 (FPD6) interaction.}
\label{tab:48cr} 
\begin{ruledtabular}
\begin{tabular}{cccccc}
Yrast state & Theor. energy & Exp. energy & $B(E2)$ $\downarrow$  & $g$-factor  & Quadrupole moment \\
\hline 
2$_{1}^{+}$  & 0.8837 (0.789) & 0.752 & 243.9 (312.4)  & 0.522 (0.518)  & -30.2 (-35.4) \\
4$_{1}^{+}$  & 1.8626 (1.940) & 1.858 & 329.2 (436.0)  & 0.524 (0.520)  & -40.4 (-45.5) \\
6$_{1}^{+}$  & 3.441 (3.657) & 3.445 & 325.9 (452.2)  & 0.531 (0.524)  & -39.1 (-48.0) \\
8$_{1}^{+}$  & 5.017 (5.569) & 5.188 & 300.6 (426.5)  & 0.533 (0.528)  & -40.6 (-48.9) \\
10$_{1}^{+}$  & 6.719 (7.664) & 7.063 & 204.9 (341.1)  & 0.542 (0.536)  & -20.4 (-41.5) \\
12$_{1}^{+}$  & 7.9704 (9.219) & 8.411 & 160.6 (152.1)  & 0.549 (0.549)  & -2.7 (-8.0) \\
14$_{1}^{+}$  & 9.994 (11.360) & 10.280 & 125.8 (137.9)  & 0.546 (0.546)  & -5.3 (-9.4) \\
16$_{1}^{+}$  & 13.226 (14.620) & 13.309 & 62.4 (68.9)  & 0.547 (0.548)  & -8.6 (-8.7) \\
18$_{1}^{+}$  & 17.731 (19.431) &  & 0.7 (2.0)  & 0.530 (0.532)  & -31.4 (-34.0) \\
20$_{1}^{+}$  & 22.478 (24.262) &  & 3.1 (7.8)  & 0.521 (0.523)  & -44.7 (-46.7) \\
\end{tabular}
\end{ruledtabular} 
\end{table}

\section{$B(E2)$'s}

The $B(E2)$'s are shown in Tables~\ref{tab:44ti} to \ref{tab:96cd}. 
To make the comparison easy, we note that for the rotational model the 
ratio $B(E2,J+2\rightarrow J)/B(E2,J\rightarrow J-2)$ for $J=0,2,4,6,8,10$ 
are respectively 1, 1.428, 1.105, 1.044, 1.027, and 1.018. The corresponding 
values in the vibrational model are 1, 2, 3, 4, 5, and 6. A rough common 
feature of all the nuclei here considered is that in the shell model (with both sets of
interactions) the ratio $B(E2,4\rightarrow 2)$ is greater than $B(E2,2\rightarrow 0)$.
This is in qualitative, if not quantitative, agreement with the two collective
models. However, with the exception of $^{88}$Ru, there is a slight
decrease in $B(E2,6\rightarrow 4)$ relative to $4\rightarrow 2$. This
is in quantitative disagreement with the collective models, although
the disagreement with the rotational model is less severe. In the $f$-$p$
shell, we then found a rapid drop-off in $B(E2)$ with increasing $J$, in
disagreement with the collective models. This is probably true for
the heavier nuclei as well, but is somewhat obscured by the fact
that we have a $J=12$ cutoff for these nuclei. The nucleus $^{88}$Ru
is unusual in that the $B(E2)$ values increase as the angular momentum
increases, more in line with a collective model picture than any other
nucleus.

For the calculated $B(E2)$'s shown in Tables~\ref{tab:44ti}-\ref{tab:96cd}, 
we find that in the lighter $f_{7/2}$ nuclei the trends are reproduced using 
either interaction. In $^{44}$Ti, the value of the $B(E2)$ increases with 
$4\rightarrow 2$, being larger than $2\rightarrow 0$, in line with the expectations
of either collective model but then decreases. In $^{48}$Cr, the
value increases again in $4\rightarrow 2$ compared to $2\rightarrow 0$,
but then remains relatively constant in the $6\rightarrow 4$ and $8\rightarrow 6$
cases before decreasing. In the case of $^{52}$Fe, the two interactions
start to show different behaviors, the $B(E2,8\rightarrow 6)$ behaving
very differently depending on which interaction we consider. 
The experimental values of the $B(E2)$'s in $^{44}$Ti, $^{48}$Cr, and $^{52}$Fe
are respectively 130, 272, and 164 e$^{2}$ fm$^{4}$.

The $^{92}$Pd calculations show a relatively flat value while the ones for
$^{96}$Cd are more like the $f{_{7}/2}$ results, where the $4\rightarrow 2$
value represents an increase over the $2\rightarrow 0$ value, but then
it immediately decreases when we look at the $6\rightarrow 4$ value
and others as we increase in angular momentum and energy. The $^{92}$Pd 
results agree with those in Refs.~\cite{cederwall11,qi11}.

Another point of interest is how the values of $B(E2)$ vary with the
number of valence particles (holes). With the first interaction in
each list, the values of $B(E2,2\rightarrow 0)$ for 4,8, and 12 valence
particles in the $f$-$p$ shell ($^{44}$Ti, $^{48}$Cr, and $^{52}$Fe)
are respecively 103, 244, and 218 e$^{2}$ fm$^{4}$. The $^{48}$Cr
value is somewhat more than a factor of 2 greater than the one for $^{44}$Ti. 
The drop-off for Fe can be explained by the fact that it can be regarded
as 4 holes relative to a closed $f_{7/2}$ shell, $Z=28$, $N=28$. Indeed
in the single-$j$-shell model the values of $B(E2)$ would be identical
for $^{52}$Fe and $^{44}$Ti.

The corresponding values for$^{96}$Cd, $^{92}$Pd, and $^{88}$Ru are
are respectively 152, 304, and 492 e$^{2}$ fm$^{4}$. Somewhat loosely
the $B(E2)$ is proportional to the number of valence holes relative
to $Z=50$, $N=50$. There are no experimental values at present for the
$B(E2)$'s in these nuclei.

\begin{table}[htb]
\caption{Excitation energies, $B(E2)$'s, $g$-factors, and quadrupole moments of
$^{52}$Fe using the gxpf1 (FPD6) interaction.}
\label{tab:52fe} 
\begin{ruledtabular}
\begin{tabular}{cccccc}
Yrast state & Theor. energy & Exp. energy & $B(E2)$ $\downarrow$  & $g$-factor  & Quadrupole moment \\
\hline 
2$_{1}^{+}$  & 0.976 (1.003) & 0.849 & 218.5 (291.2)  & 0.515 (0.515)  & -30.5 (-33.7) \\
4$_{1}^{+}$  & 2.604 (2.749) & 2.385 & 286.0 (424.3)  & 0.523 (0.520)  & -37.5 (-38.5) \\
6$_{1}^{+}$  & 4.361 (4.662) & 4.326 & 166.0 (344.5)  & 0.538 (0.520)  & -0.6 (-14.8) \\
8$_{1}^{+}$  & 6.205 (6.488) & 6.361 & 4.7 (425.3)  & 0.522 (0.514)  & -18.3 (-24.7) \\
10$_{1}^{+}$  & 7.073 (7.715) & 7.382 & 42.2 (8.7)  & 0.549 (0.553)  & 20.0 (21.3)\\
12$_{1}^{+}$  & 7.089 (8.202) & 6.958 & 57.4 (52.4)  & 0.554 (0.556)  & 54.1 (62.2) \\
14$_{1}^{+}$  & 10.920 (11.482) &  & 29.1 (34.4)  & 0.550 (0.550)  & 62.2 (64.8) \\
16$_{1}^{+}$  & 14.960 (15.777) &  & 10.7 (3.6)  & 0.536 (0.538)  & 22.7 (27.4) \\
18$_{1}^{+}$  & 19.150 (20.553) &  & 8.3 (27.8)  & 0.550 (0.536)  & 16.7 (24.4) \\
20$_{1}^{+}$  & 22.951 (23.692) &  & 2.5 (22.7)  & 0.524 (0.527)  & -8.7 (-5.4) \\
\end{tabular}
\end{ruledtabular} 
\end{table}

\begin{table}[htb]
\caption{Excitation energies, $B(E2)$'s, $g$-factors, and quadrupole moments of
$^{88}$Ru using the jun45 (jj4b) interaction.}
\label{tab:88ru} 
\begin{ruledtabular}
\begin{tabular}{cccccc}
Yrast state & Theor. energy & Exp. energy & $B(E2)$ $\downarrow$  & $g$-factor  & Quadrupole moment \\
\hline 
2$_{1}^{+}$  & 0.576 (0.566) & 0.616 & 492.0 (578.3)  &  & 36.7 (29.0) \\
4$_{1}^{+}$  & 1.314 (1.281) & 1.416 & 764.1 (842.6)  &  & 43.2 (37.1) \\
6$_{1}^{+}$  & 2.115 (2.030) & 2.380 & 890.9 (972.0)  &  & 47.5 (45.5) \\
8$_{1}^{+}$  & 2.881 (2.803) & 3.480 & 979.9 (1056.1)  &  & 52.3 (49.5) \\
10$_{1}^{+}$  & 3.674 (3.648) &  & 1061.1 (1102.4)  &  & 52.4 (51.1) \\
\end{tabular}
\end{ruledtabular} 
\end{table}

\section{Quadrupole Moments}

By looking only at $B(E2)$'s, one cannot tell if a ground state band
is prolate or oblate. For this reason we have extended the calculations
to static quadruple moments. Perhaps the most interesting result is
that for $^{88}$Ru we get a robust oblate deformation. This has already
been reported in Ref.~\cite{lz13}. We can compare this ``8-particle system''
with a corresponding one in the $f$-$p$ shell---$^{48}$ Cr. The value
of $Q(2^+)$ for $^{88}Ru$ is $+36.7$ e~fm$^{2}$, whereas it is $-30.2$ e~fm$^{2}$
for $^{48}$Cr, i.e. similar magnitudes but opposite signs. One word of
caution, the calculation for $^{88}$Ru is in a less complete model
space with only $g_{9/2}$ from the $s$-$d$-$g$ shell included. Also it
is better stated that $^{88}$Ru is a 2-hole system relative to $^{100}$Sn.

The values of $Q(2^+)$ for the ``8-hole system'' $^{92}$Pd are almost
equal and opposite for the two interactions used, $-3.5$ and $+4.6$ for
June45 and jj4b respectively. But the key point is that both are very
small. Recall that in the harmonic vibrational model $Q(2^+)$ is equal
to zero. This supports the statements in Refs.~\cite{cederwall11,qi11} about
the equally spaced levels. However the ratio of $B(E2)$'s $6\to 4/4\to 2$ would
be 1.5 in the vibrational model,whereas we calculate this ratio to
be slightly less than one with both interactions. So the entire situation
is more complicated.

Note that in $^{48}$Cr there is a dramatic drop in the magnitude
of the static quadrupole moment $Q$ when one goes from $10^+$ to $12^+$, 
from $-41.5$ to $-8.0$ e~fm$^{2}$. Similar behavior was commented in the context
of $^{50}$Cr by Zamick, Fayache, and Zheng~\cite{lz96}. They asserted that
in the rotational model the $J=10^{+}$ state of $^{50}$Cr does not
belong to the $K=0$ ground state band. Indeed it could belong to a $K=10^{+}$
band. They used static quadrupole calculations to support their claim. This
was also discussed by a dominantly experimental group, Brandolini et
al.~\cite{brand02}. 

There are no experimental values for the static quadrupole moments
of any of the nuclei here considered.

\begin{table}[htb]
\caption{Excitation energies, $B(E2)$'s, $g$-factors, and quadrupole moments of
$^{92}$Pd using the jun45 (jj4b) interaction.}
\label{tab:92pd} 
\begin{ruledtabular}
\begin{tabular}{cccccc}
Yrast state & Theor. energy & Exp. energy & $B(E2)$ $\downarrow$ & $g$-factor & Quadrupole moment \\
\hline 
2$_{1}^{+}$  & 0.840 (0.785) & 0.874 & 304.5 (366.2)  & 0.537 (0.529)  & -3.5 (4.6) \\
4$_{1}^{+}$  & 1.720 (1.750) & 1.786 & 382.6 (497.6)  & 0.539 (0.530)  & -8.0 (11.1) \\
6$_{1}^{+}$  & 2.515 (2.719) & 2.536 & 364.1 (465.2)  & 0.541 (0.534)  & -1.9 (23.9) \\
8$_{1}^{+}$  & 3.217 (3.570) &  & 315.1 (283.4)  & 0.541 (0.539)  & 8.3 (33.8) \\
10$_{1}^{+}$  & 4.070 (4.525) &  & 334.6 (344.6)  & 0.542 (0.539)  & 7.9 (40.0) \\
\end{tabular}
\end{ruledtabular} 
\end{table}

\begin{table}[htb]
\caption{Excitation energies, $B(E2)$'s, $g$-factors, and quadrupole moments of
$^{96}$Cd using the jun45 (jj4b) interaction. There are no known experimental energies.}
\label{tab:96cd} 
\begin{ruledtabular}
\begin{tabular}{ccccc}
Yrast state  & Theor. energy  & $B(E2)$ $\downarrow$  & $g$-factor  & Quadrupole moment \\
\hline 
2$_{1}^{+}$  & 0.901 (0.901)  & 151.9 (154.7)  & 0.541 (0.539)  & -19.3 (-16.4) \\
4$_{1}^{+}$  & 1.987 (1.964)  & 206.0 (205.7)  & 0.542 (0.540)  & -21.5 (-15.2) \\
6$_{1}^{+}$  & 3.021 (2.957)  & 191.0 (187.1)  & 0.542 (0.541)  & -10.5 (-2.4) \\
8$_{1}^{+}$  & 3.483 (3.404)  & 46.7 (71.4)  & 0.541 (0.540)  & 40.2 (37.2) \\
10$_{1}^{+}$  & 4.801 (4.789)  & 52.3 (80.9)  & 0.544 (0.537)  & 14.9 (24.0) \\
\end{tabular}
\end{ruledtabular} 
\end{table}

\section{$g$ factors}

We note that there is very little variation in the values of the $g$
factors. A typical value is 0.54 with small fluctuations around this
value. This result is not unexpected. In the single-$j$-shell model the
$g$ factor of any $N=Z$ even-even nucleus is given by $g=(g_{j\pi} +
g_{j\nu})/2$ for all nuclei and is independent of the details of
the wave function. In the $f_{7/2}$ shell we get $g=0.55$. In the $g_{9/2}$
shell we get...

Additionally, this is very close to the collective $Z/A$ value of
0.5. Either extreme picture of pure collectivity and pure single $j$-shell
yields values close to this value. This has previously been commented
upon by Yeager et al.~\cite{sy09}.

The experimental value of the $g$ factor of $^{44}$Ti is 0.5(.15).
No other factors in this work have been measured.

\section{Comparison of shell model occupancies with different interactions.}

In this section we point out that there are surprising differences
in the occupation percentages that result when different ``standard''
interactions are used. The importance of getting correct occupancies
via transfer reactions, e.g. (d,p) and (p,d), has been emphasized over
the years by John Schiffer and collaborators. We here cite only the
most recent, Ref.~\cite{kay13}.

In Table~\ref{tab:occu} we give the percent occupancy of the lowest configuration
of the $J=0^{+}$ ground state and first $2^{+}$ state. By this we
mean the percent occupancy of the state (4,6,2,8) in $^{88}Ru$ and
(4,6,2,12) in $^{92}$Pd. We also give $B(E2,2\rightarrow 0)$ and
$Q(2^{+})$ for the two interactions. We see that interactions which
have lower occupancies of the lowest states (i.e. more fragmentation)
have larger $B(E2)$'s. The situation with the static quadrupole moments
is more complicated. As mentioned before, the $Q(2^+)$ values for $^{92}$Pd
are small and of opposite sign, possibly indicating vibrational behavior.
And most surprising, with both interactions the values of $Q(2^+)$ are
large and positive for $^{88}$Ru, an indication of an oblate deformation.
Whether this result persists when larger model spaces become feasible
remains to be seen. 

Also of interest is the fact that the $B(E2)$'s increase with the number
of valence particles almost in a linear fashion, e.g. 155, 366, and
579 e$^{2}$ fm$^{4}$ for $A=96$, 92, and 88 (4, 8, and 12 holes relative
to the doubly closed shell $^{100}$Sn). We also have here noted dramatic
changes in static quadrupole moments $Q(J)$ beyond certain spin values, an
indication perhaps of changing from $K=0$ bands to high-$K$ bands. This is
certainly worthy of future study.

In closing we note that, although the collective models can supply
valuable insights concerning the behaviors of electromagnetic properties
of nuclei, the simplest versions of these models are clearly inadequate.
For example they fail to predict the decrease in $B(E2)$'s after a certain
point with increasing spin. Undoubtedly more sophisticated collective
models can be constructed which might be more successful, but then the simplicity
is lost and the insights obscured. 

The large scale shell models are not off the hook either. One must
remember that they depend on what interactions are used and the current
state of affairs is such that different widely used interactions can
and do yield quite different results, and these can be most easily
traced to the occupation numbers for various basis states. Also the
model spaces may be too restricted. For the heavier nuclei the orbits
included are $p_{3/2}$, $f_{5/2}$, $p_{1/2}$, and $g_{9/2}$. It would
be good to have more positive-parity orbits.

We conclude by noting that the region below $^{100}$Sn has been very
active of late. Besides the references already mentioned, we add 
Refs.~\cite{qi10}-\cite{lz13b}. Also, to a large extent, one can regard earlier 
studies of properties in the $f_{7/2}$ shell as precursors to analogous studies 
in the $g_{9/2}$ shell. This has been made especially clear by Neergaard~\cite{kn13}. 
There is a marked similarity in the structure of ground state wave functions 
in the two shells~\cite{bfb64,jg63}. In the last reference, Girod~\cite{giord13} 
presents a cluster model of $^{88}$Ru consisting of four $^{16}$O and two 
$^{12}$C nuclei. It would be of interest to make a connection of this with our 
oblate $^{88}$Ru .

\begin{table}[htb]
\caption{Occupation percentages for different interactions.}
\label{tab:occu} 
\begin{ruledtabular}
\begin{tabular}{cccccc}
Nucleus & Interaction & $J=0^{+}$ & $J=2^{+}$ & $B(E2)$ & $Q(2^+)$ \\
\hline
$^{44}$Ti & gx1 & 72.1 & 66.1 & 103 & $-5.1$ \\
 & fpd6 & 42.9 & 26.8 & 140 & $-21.6$ \\ \hline
$^{48}$Cr & gx1 & 43.2 & 34.7 & 244 & $-30.1$ \\
 & fpd6 & 21.2 & 16.2 & 312 & $-35.4$ \\ \hline
$^{96}$Cd & jj44b & 49.6 & 61.0 & 155 & $-16.4$ \\
 & jun45 & 58.8 & 76.4 & 152 & $-19.3$ \\ \hline
$^{92}$Pd & jj44b & 9.7 & 9.0 & 366 & 4.6 \\
 & jun45 & 28.8 & 32.6 & 304 & $-3.5$ \\ \hline
$^{88}$Ru & jj44b & 1.65 & 1.24 & 578 & 29.0 \\
 & jun45 & 7.14 & 5.29 & 492 & 36.7 \\
\end{tabular}
\end{ruledtabular} 
\end{table}


\begin{thebibliography}{References}
\bibitem{sjqr05} S.J.Q. Robinson, A. Escuderos, and L. Zamick,  Phys. Rev.
C72, 034314 (2005).

\bibitem{cederwall11} B. Cederwall et al., Nature (London) 469, 68 (2011).

\bibitem{qi11} C.Qi, J. Blomqvist, T. Back, B. Cederwall, A. Johnson,
R.J. Liotta, and R. Wyss, Phys.Rev. C84, 021301 (R) (2011).

\bibitem{lz13} L. Zamick, S.J.Q. Robinson, T. Hoang, Y.Y. Sharon, and
A. Escuderos, BAPS, 2013 Fall Meeting of the APS Division of Nuclear
Physics, vol. 58, number 13, EA.00174 (Oct. 2013).

\bibitem{honma09} M. Honma, T. Otsuka, T. Mitzusaki, and M. Hjorth-Jensen, Phys.
Rev. C80, 064323 (2009).

\bibitem{sjqr11} S.J.Q. Robinson, L. Zamick, and Y.Y. Sharon, Phys.Rev. C83, 027302 (2011).

\bibitem{lz96} L.Zamick and D.C. Zheng, Phys. Rev. C54, 956 (1996).

\bibitem{brand02} F. Brandolini et al., Phys. Rev. C66, 021302(R) (2002).

\bibitem{sy09} S. Yeager, L. Zamick, Y.Y. Sharon, and S.J.Q. Robinson,
Europhys. Lett. 88, 52001 (2009).

\bibitem{kay13} B.P. Kay, J.P. Schiffer, and S.J. Freeman, Phys. Rev.
Lett. 111, 042502 (2013).

\bibitem{qi10} C.Qi, Phys. Rev. C81, 0343018 (2010).

\bibitem{sz11} S. Zerguine and P. Van Isacker, Phys.Rev. C83, 064314
(2011).

\bibitem{2qi11} C.Qi, R.J. Liotta, and R. Wyss, Proceedings Conf. Advanced
Many Body Systems and Statistical Methods in Microscopic Systems,
J. Phys: Conf. Ser. 338, 012207 (2011).

\bibitem{corr12} L. Corragio, A. Covello, A. Gargano, and N. Itaco, Phys.
Rev. C 85, 034335 (2012).

\bibitem{xu12} Z.X. Xu, C. Qi, J. Blomqvist, R.J. Liotta, and R. Wyss,
Nucl. Phys. A877, 51 (2012).

\bibitem{lz12} L. Zamick and A. Escuderos, Nucl. Phys. A 889, 8 (2012).

\bibitem{lz13b} L. Zamick and A. Escuderos, Phys. Rev. C87, 044302 (2013).

\bibitem{kn13} K. Neergaard, Phys. Rev. C 88, 034329 (2013).

\bibitem{bfb64} B.F. Bayman, J.D. McCullen, and L. Zamick, Phys. Rev.
134, B515 (1964).

\bibitem{jg63} J. Ginocchio and J.B. French, Phys. Lett. 7, 137 (1963).

\bibitem{giord13} M. Girod, Clustering in Nuclei with the Gogny Interaction,
ESTN May 30-31, 1 (2013).
\end{thebibliography}
\end{document}